\documentclass{article}

\usepackage{PRIMEarxiv}

\usepackage[utf8]{inputenc} % allow utf-8 input
\usepackage[T1]{fontenc}    % use 8-bit T1 fonts
\usepackage{hyperref}       % hyperlinks
\usepackage{url}            % simple URL typesetting
\usepackage{booktabs}       % professional-quality tables
\usepackage{amsfonts}       % blackboard math symbols
\usepackage{nicefrac}       % compact symbols for 1/2, etc.
\usepackage{microtype}      % microtypography
\usepackage{lipsum}
\usepackage{fancyhdr}       % header
\usepackage{graphicx}       % graphics
\usepackage{subcaption}
\graphicspath{{media/}}    
\usepackage{amssymb}
\usepackage{mathrsfs}
\usepackage{amsmath}
\usepackage{amsfonts}
\usepackage{graphicx}
\usepackage{xcolor}
\usepackage{graphicx}
\usepackage{bm} %bold math
\usepackage{footnote}
\usepackage{comment}
\usepackage{numprint}
\usepackage{siunitx}
\usepackage{booktabs} 
  
\title{Application of probabilistic modeling and automated machine learning framework for high-dimensional stress field}

\author{
Lele Luan\\
Probabilistic Design Laboratory\\
General Electric Research\\
Niskayuna, NY 12309\\	
\And
Nesar Ramachandra\\
Computational Science Division \\
Argonne National Laboratory\\
Lemont, IL 60439\\
\And
Sandipp Krishnan Ravi\\
Probabilistic Design Laboratory\\
General Electric Research\\
Niskayuna, NY 12309\\	
\And
Anindya Bhaduri\\
Probabilistic Design Laboratory\\
General Electric Research\\
Niskayuna, NY 12309\\
\And
Piyush Pandita
\thanks{piyush.pandita@ge.com}\\
Probabilistic Design Laboratory\\
General Electric Research\\
Niskayuna, NY 12309\\	
\And
Prasanna Balaprakash\\
Oak Ridge National Laboratory \\
1 Bethel Valley Rd, Oak Ridge, TN\\	
\And
Mihai Anitescu\\
Mathematics and Computer Science Division \\
Argonne National Laboratory\\
Lemont, IL 60439\\	
\And
Changjie Sun\\
Materials and Mechanical Systems\\
General Electric Research\\
Niskayuna, NY 12309\\
\And
Liping Wang\\
Probabilistic Design Laboratory\\
General Electric Research\\
Niskayuna, NY 12309\\	
}

\begin{document}
\maketitle
\begin{abstract}
Modern computational methods involving highly sophisticated mathematical formulations enable several tasks like modeling complex physical phenomena, predicting key properties, and optimizing design.
The higher fidelity in these computer models makes it computationally intensive to query them hundreds of times for optimization. One usually relies on a simplified model, albeit at the cost of losing predictive accuracy and precision.
Towards this, data-driven surrogate modeling methods have shown much promise in emulating the behavior of expensive computer models.
However, a major bottleneck in such methods is the  inability to deal with high input dimensionality and the need for relatively large datasets. 
In certain cases, the high dimensionality of the input space can be attributed to its image-like characteristics, for example, the stress and displacement fields of continuums. 
With such problems, the input and output quantity of interest are tensors of high dimensionality.
Commonly used surrogate modeling methods for such problems suffer from requirements like many computational evaluations that precludes one from performing other numerical tasks like uncertainty quantification and statistical analysis.
This work proposes an end-to-end approach that maps a high-dimensional image-like input to an output of high dimensionality or its key statistics.
Our approach uses two main frameworks that perform three  steps: a) reduce the input and output from a high-dimensional space to a reduced or low-dimensional space, b) model the input-output relationship in the low-dimensional space, and c) enable the incorporation of domain-specific physical constraints as masks.
To reduce input dimensionality, we leverage principal component analysis, coupled with two surrogate modeling methods: a) Bayesian hybrid modeling and b) DeepHyper's deep neural networks.
We demonstrate the approach's applicability to a linear elastic stress field data problem.
We perform numerical studies to study the effect of the two end-to-end workflows and the effect of data size.
Key insights and conclusions are provided, which can aid such efforts in surrogate modeling and engineering optimization.
\keywords{Surrogate modeling, Image-based models, Deep neural networks, Dimension reduction, Bayesian hybrid modeling}
\end{abstract}

\section {Introduction}
\label{sec:intro}

High-fidelity computer simulations~\cite{michopoulos2005modeling} provides an elegant way to predict a physical system's behavior by modeling the corresponding physical process using numerically intensive mathematical equations. 
Such high-fidelity computational models are the bedrock of everyday modeling and analysis of physical systems in research laboratories worldwide.
The elevated sophistication in the predictability of the underlying physical response usually comes at a severe computational cost.
This precludes a researcher or scientist from gaining in-depth insight into the physical problem, as performing tasks like uncertainty propagation, sensitivity analysis, optimization, and error analysis require hundreds of runs of the computational model.
This is a common situation in mechanical and aerospace engineering. Several physics processes like optimizing energy-intensive friction stir welding~\cite{mishra2011friction} of aircraft panels, aerodynamic analysis of turbomachinery components and flows through porous media~\cite{tsilifis2017reduced,pandita2016extending} require detailed analysis before allocating budget for experiments.

To enable tasks like sensitivity analysis and design optimization, using surrogate modeling methods~\cite{gunst1996response, bhaduri2018efficient, bhaduri2020free, bhaduri2020usefulness, cristianini2000introduction, williams1998prediction, bhaduri2021probabilistic, bhaduri2022stress} to build the so-called \emph{emulators} is a common practice.
Several approaches for surrogate modeling have shown promise across a myriad of engineering problems~\cite{zhao2022comparative,tran2022integrated,ravi2022data,ravi2022data_1,pei2022multi,roy2023data,ravi2023uncertainty,roy2022elucidating,roy2023understanding}. 
Commonly known methods include Gaussian process regression (GPR)~\cite{williams2006gaussian,liu2018gaussian,quinonero2005unifying}, deep neural networks (DNN), and polynomial chaos expansions (PCE)~\cite{ghanem_spanos}.
The key idea behind employing a surrogate model is its ability to accurately emulate the input-output relation by inferring physically reasonable correlations from data.
Methods like GPR and PCE have shown immense promise in modeling datasets that stem from computationally expensive solvers, hence working well under a finite budget.
Additionally, DNN-based models that are considered data-hungry have more advanced versions ~\cite{graves2011practical,paisley2012variational,deshpande2019computational} that have been noted for their excellent regression performance across various challenging applications.

The applicability of these methods is severely hampered by datasets that exhibit more granular characteristics of the underlying problem, like high dimensionality or image-like input and output structure. 
One way to overcome the challenges associated with high-dimensional models is to apply dimensionality reduction techniques~\cite{constantine_AS,constantine_book}. 
Each observation contains many dimensions or features for high-dimensional data. Dimensionality reduction techniques compress the data into a lower-dimensional manifold while preserving the maximum data information. 
As a data pre-processing step in machine learning, dimensionality reduction techniques not only help increase the overall performance of modeling by avoiding the problem of overfitting and denoising data but also enables the visualization of multidimensional data. 

In recent years, linear and nonlinear dimensionality reduction methods have been proposed to incorporate surrogates for high-dimensional data modeling. 
The linear dimensionality reduction methods used for surrogate modeling building include Principal Component Analysis (PCA)~\cite{hombal2013surrogate}, proper orthogonal decomposition (POD) \cite{zimmermann2013gradient} and gradient-free Bayesian methods~\cite{tripathy2016gaussian}. 
Nonlinear dimension reduction methods, like variational auto-encoders, manifold learning methods~\cite{soize2022probabilistic} and gradient information-based nonlinear dimension reduction~\cite{bigoni2021nonlinear} are used to circumvent the assumptions in the linear methods that high-dimensional data can only be compressed in linear spaces. 
 
In this work, we apply two high-dimensional data modeling approaches, augmented with a dimension reduction technique, on image-like datasets generated from an expensive computational solver. 
In both approaches, the high-dimensional data is first reduced to low-dimensional vectors and then mapped to the high-dimensional output image using two separate modeling methods, namely, a) Bayesian hybrid models and b) deep neural networks with automated hyperparameter tuning and neural architecture search.
The first step involves linear dimensionality reduction using principal component analysis (PCA), a commonly used method for compressing high-dimensional input space to a finite number of independent components.
A key feature of the two frameworks is the automated nature of the dimension reduction that is seamlessly blended with the two forward modeling methods without manual tuning.
The Bayesian surrogate modeling is done using GE's in-house Bayesian hybrid modeling (GE-BHM)~\cite{ghosh2020advances} suite of methods, and the automated deep neural network-based models are asynchronously trained using DeepHyper~\cite{balaprakash2018deephyper}.
We demonstrate the performance of the two approaches on an engineering problem of modeling linear elastic stress field maps as a function of the input shape of the ellipsoidal void in a uniform continuum. 
The problem is set up using ANSYS APDL scripts and automated through Python scripts. 
The problem setup is a controllable and versatile representation of high computational simulations.
The rest of the paper is organized as follows: first, we describe the engineering problem in Section \ref{sec:problem_description}, next we provide an overview of the different modeling methods used for dimension reduction and surrogate modeling in Section \ref{sec:methodology}, followed by the numerical results on the engineering problem of high-dimensional elastic stress field modeling in Sec.\ref{sec:results}. 
We summarize our conclusions from this work in Section \ref{sec:conclusions}.

\section{Data description }
\label{sec:problem_description}
To understand and investigate the framework proposed in this work, we formulate a suitable simulation with an image-like input and output data structure.
With the need for spatial data in congruence with mechanical relevance, the problem is conceived as a solid cuboid incorporated with an ellipsoidal cavity subjected to uniaxial load. 
The dimensions of the cuboidal domain and the void are decided based on the requirement of \emph{far field} stress conditions to prevent any edge effects. 

\begin{figure}[t!]
\centering
\includegraphics[width=0.3\textwidth]{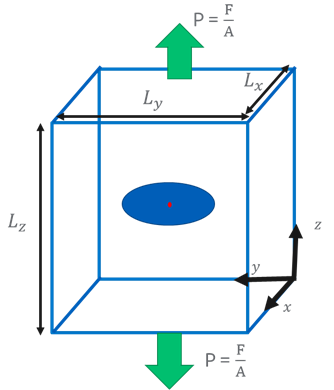}
\caption{Schematic of the ellipsoidal void Problem}
\label{fig:Ellip_Prob}
\end{figure}

A schematic of the problem is given in Figure - (\ref{fig:Ellip_Prob}). 
The center of the ellipsoid coincides with the center of the cuboid. 
The dimensions of the cuboid are all set to 0.1 m ($L_x = L_y = L_z = 0.1m$). 
The different cases of simulations are generated by varying the axis lengths of the ellipsoidal void ($R_x,R_y,R_z$) and angles of rotation about the central axis ($\theta_x,\theta_y,\theta_z$). 
Based on the existence of rotation, two categories of datasets are generated. 
In the rest of this paper, these two cases are named as follows: a) Non-Rotated Ellipse Case, b) Rotated Ellipse Case. 
The bounds of the ellipsoid axis length and angle variations are defined in Tables (\ref{tab:table_nonrot} - \ref{tab:table_rot}).

\begin{table}
\caption{\label{tab:table_nonrot} Summary of Non-Rotated Ellipse Case}
\centering
\begin{tabular}{lccc}
\hline
Variable & Minimum  & Maximum \\
\hline
$R_x$ & $0.05 \times L_x$ & $0.1 \times L_x$  \\
$R_y$ & $0.05 \times L_y$ & $0.1 \times L_y$  \\
$R_z$ & $0.05 \times L_z$ & $0.1 \times L_z$  \\
$\theta_x$ & 0 & 0  \\
$\theta_y$ & 0 & 0 \\
$\theta_z$ & 0 & 0 \\
\hline
\end{tabular}
\end{table}

\begin{table}
\caption{\label{tab:table_rot} Summary of Rotated Ellipse Case}
\centering
\begin{tabular}{lccc}
\hline
Variable & Minimum  & Maximum \\
\hline
$R_x$ & $0.05 \times L_x$ & $0.1 \times L_x$  \\
$R_y$ & $0.05 \times L_y$ & $0.1 \times L_y$  \\
$R_z$ & $0.05 \times L_z$ & $0.1 \times L_z$  \\
$\theta_x$ & 0 & 0  \\
$\theta_y$ & $\frac{\pi}{36}\ $ & $\frac{17\pi}{36}\ $ \\
$\theta_z$ & 0 & 0 \\
\hline
\end{tabular}
\end{table}

\begin{figure}[t!]
\centering
\includegraphics[width=0.2\textwidth]{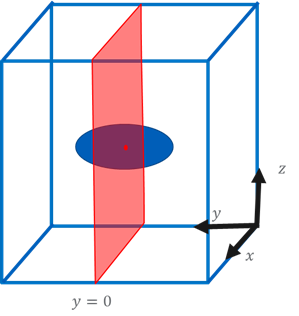}
\caption{Schematic of 2D Approximation}
\label{fig:Ellip_Prob_2D}
\end{figure}

\begin{figure}[t!]
\centering
\includegraphics[width=0.5\textwidth]{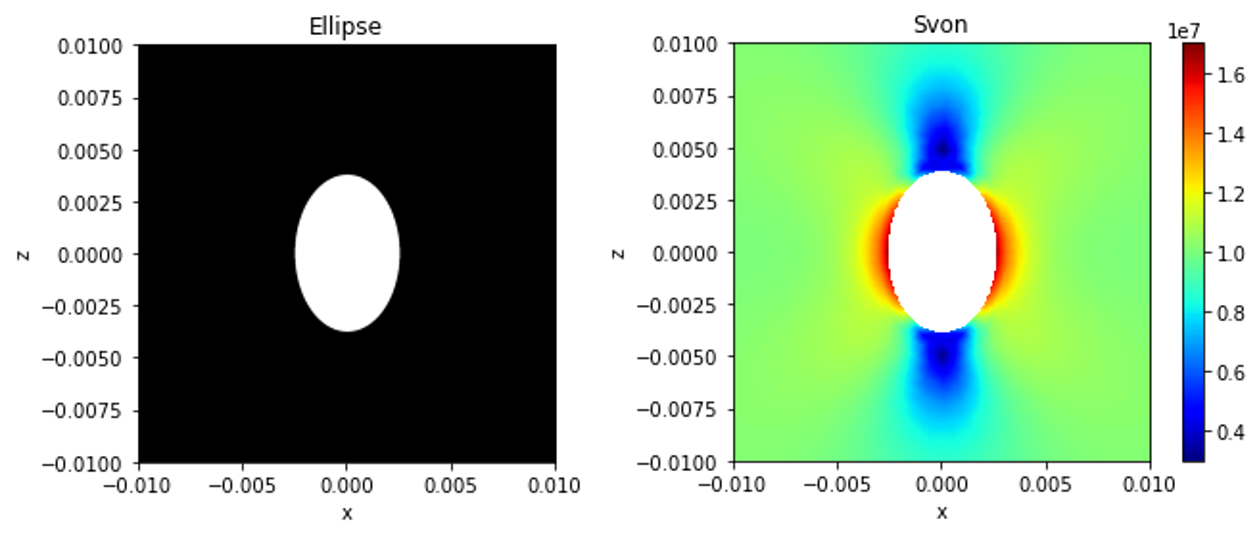}
\caption{Representative Example of Non-Rotated Cases}
\label{fig:Ellip_Prob_Res_cse1}
\end{figure}

\begin{figure}[t!]
\centering
\includegraphics[width=0.5\textwidth]{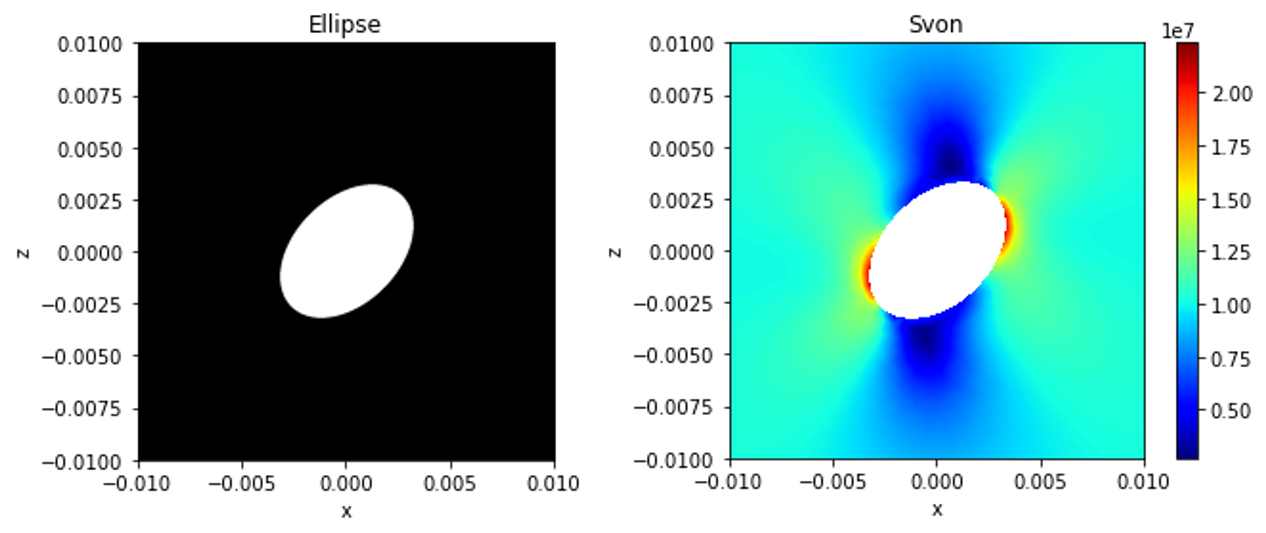}
\caption{Representative Example of Rotated Cases}
\label{fig:Ellip_Prob_Res_cse1_rot}
\end{figure}

A uniaxial uniformly distributed point load is applied on the top and bottom edge of the cuboid domain. 
The material properties and other specifications of the simulations are given through Table - \ref{tab:table_matprop}. 
The entire data generation is automated by leveraging  scripts in the Python programming language  coupled with ANSYS APDL scripts. 
Each simulation run takes approximately 15 min. 
250 data points are generated with randomly sampled ellipsoidal axis lengths for non-rotated cases. 
On the other hand, a total of 250 data points are generated with randomly sampled ellipsoidal axis lengths and angles for rotated cases. 
The outputs of interest are the nodal values of the Von Mises stress. 
%The distribution of parameters for the two cases are given through Figure - (\ref{fig:Ellip_Prob_Dist}-\ref{fig:Ellip_Prob_Dist_rot}) to get a better insight into the design space explored. 
Even though the data originally generated is three-dimensional, the surrogate modeling and dimensionality reduction is set up for a two-dimensional setting to enable faster development cycles. 
To accommodate this, the three-dimensional stress field is converted to a two-dimensional stress field by cutting a XZ plane at Y = 0 as shown in Figure - (\ref{fig:Ellip_Prob_2D}). 
Representative results from the simulation for the two categories of the problems is presented through Figure - (\ref{fig:Ellip_Prob_Res_cse1}-\ref{fig:Ellip_Prob_Res_cse1_rot}). 
All the subsequent surrogate modeling tasks presented in the following sections are built upon leveraging the two-dimensional dataset.

\begin{table}
\caption{\label{tab:table_matprop} Material Properties for Ellipsoidal Void Problem}
\centering
\begin{tabular}{lccc}
\hline
Property & Value \\
\hline
$Young's \ \ Modulus \ (E)$ & $200 \times 10^{9} \: Pa$   \\
$Load \: (F) $ & $5 \times 10^{7} \: N$   \\
\hline
\end{tabular}
\end{table}

%\begin{figure}[t!]
%\centering
%\includegraphics[width=0.3\textwidth]%{figures/Ellip_Prob_Dist.png}
%\caption{Distribution of $Rx,Ry,Rz$ for Non-Rotated Cases}
%\label{fig:Ellip_Prob_Dist}
%\end{figure}

%\begin{figure}[t!]
%\centering
%\includegraphics[width=0.3\textwidth]%{figures/Ellip_Prob_Dist.png}
%\caption{Distribution of $Rx,Ry,Rz$ for Rotated Cases}
%\label{fig:Ellip_Prob_Dist_rot}
%\end{figure}

\section{Methodology} \label{sec:methodology}

We choose a modular pipeline involving latent space mapping to build the mapping from void shapes to the von Mises fields. 
A dimensionality reduction algorithm is deployed on input shapes and output von Mises fields. A supervised machine learning framework is trained to understand the relationship between the latent space variables. 

\begin{figure}[t!]
\centering
\includegraphics[width=0.45\textwidth]{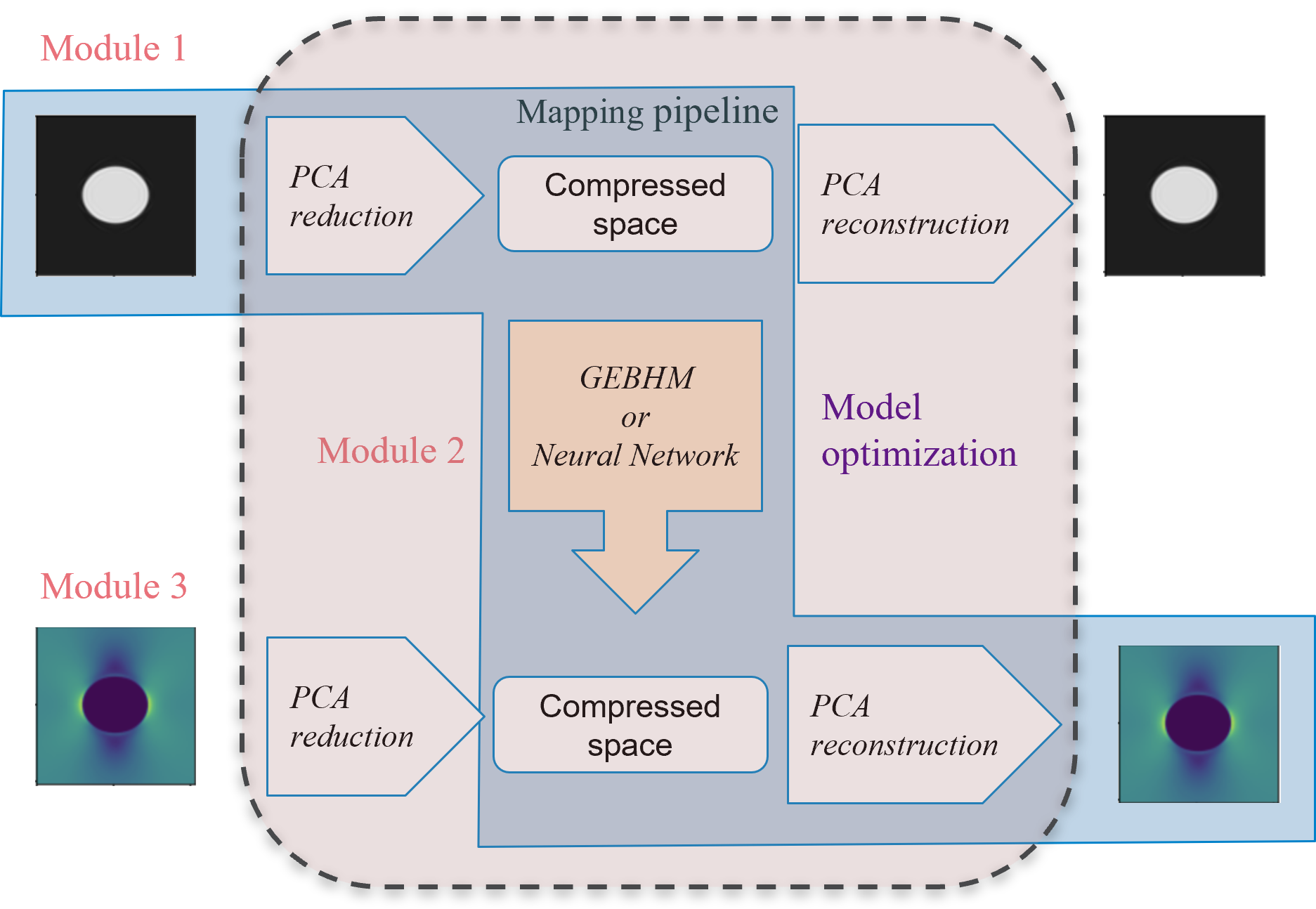}
\caption{Workflow of dimensionality reduction modeling. 
The high-dimensional input shape and output response are compressed into low-dimensional space by two PCA modules at the top and the bottom. 
The third module is a GEBHM or neural network mapping between compressed fields of both the PCA.
The red-shaded region is the modeling learning and optimization. The blue-shaded region shows the workflow of the whole model prediction.}
\label{fig:dim_red_mod}
\end{figure}

This section describes two frameworks that implement the dimensionality reduction-based algorithms for the problem described above. 
We begin with the description of the individual components of the modular framework. 
This includes transforming high-dimensional data (input shapes and output stress fields) into the low-dimensional representation by dimensionality reduction techniques and the mapping routines between the low-dimensional representations. 
Then we explain the Bayesian Hybrid Modeling (BHM) or DeepHyper-optimized deep neural network (DNN) employed to model the input-output relationship in the low-dimensional space. 
Fig.~\ref{fig:dim_red_mod} shows our
dimensionality reduction-based modeling workflow. 

\subsection{Modular components of the framework}
\label{sec:base}

There are 3 individual modules in both the workflows shown in this work. The first and the last modules are the dimension reductions to compress the high-dimensional inputs and outputs. The central module is a model to map between the compressed latent space variables. Details on the dimension reduction, the two surrogate modeling methods, and the optimization routines are given in this section.

\subsubsection{Data compression using Principal Component Analysis}
\label{sec:pca}

In dimensionality reduction, the high dimensional input and output pairs are compressed into lower dimensional vector space (the latent space). The compressed low-dimensional data enables the applications of surrogate modeling or neural networks. Hence such transformations are widely used as data pre-processing techniques to reduce the number of features in machine learning tasks. The most popular dimensionality reduction methods~\cite{sorzano2014survey} include the Principal Component Analysis (PCA), Kernel Principal Component Analysis (KPCA), t-distributed Stochastic Neighbor Embedding (t-SNE), Autoencoders, etc. In this work, we resort to the linear PCA transformation and proceed with the forward surrogate modeling using GEBHM and DeepHyper's DNN capabilities.

PCA is an orthogonal linear transformation to compress the high-dimensional data into its low-dimensional representatives~\cite{abdi2010principal}. 
PCA transforms the dimensionality of the data in the physical space to a reduced set of independent dimensions called components. Since the PCA transform is linear and operates directly on the covariance matrix of the data structure, the reduced number of components can be filtered using a criterion based on a cutoff on the percent of the variation in the data distribution captured by a said number of components. This ``truncated'' principal component analysis can reduce the computational cost of training machine learning models and improve their performance. This is particularly important when the small size of the training samples limits the complexity of the model. 

\subsubsection{Latent space mapping using GEBHM} \label{sec:gebhm}

The first approach adopted as the central module in our workflow to map the connections between the low-dimensional input and output is the GE Bayesian Hybrid Modeling (BHM)~\cite{ghosh2020advances}. 
 
\begin{figure}[t]
	\begin{center}
		\includegraphics[scale=0.55]{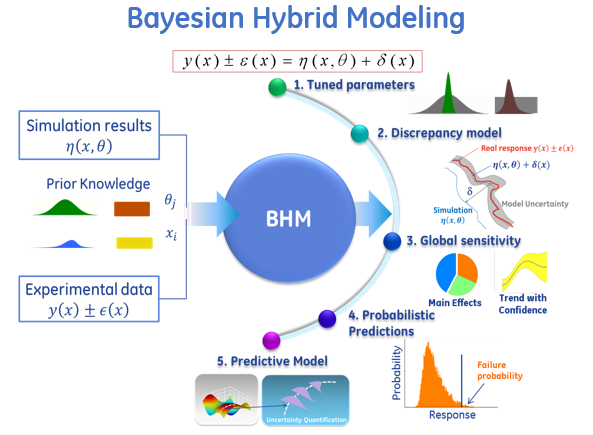}
	\end{center}
	\caption{The GE Bayesian Hybrid Modeling (GEBHM) framework: the in-house implementation of the Kennedy O'Hagan framework.}
	\label{fig:BHM} 
\end{figure}

The GEBHM framework is based on  the Kennedy \& O'Hagan \cite{kennedy2001bayesian} approach to accomplish a variety of tasks such as building surrogate models, performing probabilistic calibration of simulation parameters with respect to observed data, and building a metamodel for the discrepancy between a calibrated simulation model and observed data. 
GEBHM and GEBHM-based optimization have been tested or applied in many real engineering problems \cite{wang2011challenges, subramaniyan2012enhancing, kumar2012improving, kumar2013calibrating, ling2018intelligent, kristensen2016expected}. Fig. \ref{fig:BHM} shows a general overview of the GEBHM framework.  

In this work, the function of surrogate modeling in GEBHM is leveraged for input and output latent space mapping. We employ the fully Bayesian Gaussian process regression~\cite{williams2006gaussian} capability in the GEBHM. The details of the general approach and the specific GEBHM formulation have been discussed in~\cite{kumar2013calibrating}. Our objective is to build a probabilistic model using limited data, using input-output pairs in the low-dimensional space, and obtain an inexpensive surrogate model that can make predictions at unseen input points.
This model is then validated for predictive accuracy on held-out test data. Once validated, the model facilitates several downstream tasks for the image-to-image ellipsoidal problems, including uncertainty analysis and sensitivity studies.

\subsubsection{Latent space mapping using Fully connected neural network}
\label{sec:nn}

Following the workflow shown in Fig.~\ref{fig:dim_red_mod}, we propose an alternative method from Section \ref{sec:gebhm}. We make use of a fully connected neural network for the latent space modeling, as opposed to using a Gaussian-Process-based GEBHM modeling. Before the modeling with a neural network, two PCA-based compression modules are learned by following the same steps as for GEBHM modeling.

This neural network maps the geometries' and stress field outputs' compressed spaces. 
The fully connected dense network is parameterized with $N_{depth}$ layers and the number of neurons per layer (parametrized by the maximum width $N_{width}$) as: $[k_{1} (\textrm{input})\rightarrow N_{width} \rightarrow N_{width}/2 \rightarrow \dotsc \rightarrow N_{width}/2 \rightarrow N_{width} \rightarrow k_{2} (\textrm{output})]$. 
Each layer has a non-linear activation function, the choice of which is iterated during optimization. 
In addition, the learning rate, decay rate, batch sizes, $N_{width}$, and $N_{depth}$ are also free hyper-parameters collectively denoted as $\theta$ that are required to be specified or optimized.

The latent representation of the input geometry undergoes a forward pass through the neural network and the output is computed. The loss is computed by as the mean-square-error between the forward-passed values and the latent space representation of the von Mises fields. This loss is utilized in optimizing the neural network weights using the feedback loop of backward propagation. 

\subsection{Analysis Pipeline and Optimization}
\label{sec:pipe}

With the modules mentioned in Section \ref{sec:base}, we now create two frameworks for mapping the void geometries to the von Mises stress fields. From the two learned PCA-based modules that support the compression and reconstruction of input void shape and output stress field, only the compressive part of the PCA algorithm that maps from the input shapes to the latent space vectors (in the first module) is used in the first module. Similarly, only the reconstructive part of the PCA algorithm that maps the stress response from latent space variables is used in the last module. The compressed space of the input geometry is connected to the compressed space of the von Mises stress fields using surrogate modeling.

This section provides the details of the two frameworks implemented in this work. Apart from the central module being different between the two approaches, we also have 2 different ways of optimizing the mapping algorithms and the two PCA models.

\subsubsection{Framework 1: PCA-GEBHM-PCA}
\label{sec:bhm_mod}

A detailed description of how the different modules combine to realize the method proposed in Fig.~\ref{fig:dim_red_mod} is given in this section.

As previously mentioned, the first and the third modules with the PCA algorithm (from Section \ref{sec:pca}) compress the high-dimensional input shapes and output stress fields into the low-dimensional latent representations. The connection between the input and output low-dimensional representations is learned through the second module of GEBHM, explained in Section \ref{sec:gebhm}. 
In both PCA-based modules, we select the top $k_1$ and $k_2$ principal components, respectively, which can be determined by setting the percent of the variation in the data samples captured by the components. In addition, since the images of input shapes and output stress fields are different, the number of components with the same percentage of variance in the data captured by PCA may be different. Since the input void images are parameterized through 1s and 0s for solid and empty fields, respectively, they are not pre-processed through normalization before compression. In contrast, normalization is done for stress field compression. 
% Once the top PCA components are chosen, only the compressive transformation part of the PCA is used to convert the images to latent variables. 
Subsequently, a trained GEBHM surrogate model connects the compressed latent vectors from input shapes and stress fields. %With these components trained on the data in a self-supervised scheme, we then develop the surrogate modeling for latent space mapping with GEBHM.

%Finally, after these two PCA-based compression modules and the GEBHM-based surrogate model being trained, this whole dimensionality reduction model is able to do the prediction. 
Once each framework module is trained, the following steps enable the prediction. A test input shape image is fed into the first module for dimension reduction. The compressed shape vector is passed to the GEBHM surrogate model to predict the latent space of its corresponding stress field. The inverse transformation obtains the final output stress field via the last module.
It should be highlighted that, to ensure that the stress values at the void are 0 throughout, a masking operation is conducted on the original high-dimensional response reconstructed from the inverse transformation of the compression model. The blue shaded region further highlights this workflow in Fig.~\ref{fig:dim_red_mod}. 

In the PCA-GEBHM-PCA model, the BHM is self-contained in terms of its hyperparameters tuning. Only the hyperparameters in compression modules, that is the number of components, need to be fine-tuned. This is done through a grid search approach, where the entire framework is trained systemically for different numbers of components for the input and output compression. The final model is selected based on performance on the validation dataset. The search space for the components is selected based on the representative variance capture (like 95\%, 97\%, etc.). To save the computation cost in the grid search, the captured percentages of sample variance for input shapes and stress field are set as the same.

\subsubsection{Framework 2: PCA-NN-PCA}

This method follows the basic flow of the dimensionality reduction modeling shown in Fig.~\ref{fig:dim_red_mod}. As an alternative method from Section \ref{sec:bhm_mod}: we make use of a fully connected Neural network (Section \ref{sec:nn}) for modeling in the latent space, as opposed to using a Gaussian-Process-based GEBHM modeling. Before the modeling with a neural network, two PCA-based compression modules are learned by following the same steps as for GEBHM modeling. 

Unlike GEBHM, the neural network we use is not self-optimized, i.e., the hyperparameters and the geometries associated with the neural network have to be configured according to the data. 
For this purpose, we use the DeepHyper, a distributed machine learning (AutoML) package for automating the development of deep neural networks for scientific applications. 
The deployment of DeepHyper serves 3 purposes: (i) the neural architecture search and hyperparameter search algorithms of DeepHyper circumvent the need for exhaustive manual tuning of the modules, (ii) the ensemble of well-performing ML-models provided by DeepHyper will be used in uncertainty quantification, instead of a single best prediction model, and (iii) it effectively utilizes the computing resources by asynchronously deploying the surrogate models on a large number of nodes. 

In this work, we optimize the entire pipeline of compression instead of just the surrogate model. That is, the neural network and the 2 PCA models are optimized simultaneously. For the PCA models, the only hyper-parameters to be optimized are the number of eigenvectors $k_1$ and $k_2$ in the truncated space. 
For the neural network, the corresponding hyperparameter choices are $\theta$. 
This includes the learning rate, decay rate, batch size, and neural architecture parameters such as activation functions, number of layers, and the maximum width of the neural network.

\section{Results}
\label{sec:results}
These two dimensionality-reduction-based frameworks are applied to model the ellipsoid datasets described in section~\ref{sec:problem_description}. 
The non-rotated and rotated dataset cases are modeled separately. 
To make the prediction performance of these two frameworks comparable, both models are trained, validated, and tested on the same data samples. In the base models for both non-rotated and rotated cases, 100 data samples are selected for training and validation, 90\% for training and 10\% for validation. An additional 150 data samples are used for testing. 
The results presented are from the best model obtained from the grid search (in the PCA-GEBHM-PCA model) or DeepHyper (in the PCA-NN-PCA approach). After the analysis on base models, the performance of the two frameworks is also studied across different sizes of training samples to understand the impact of data availability.
To enable a uniform quantification of the prediction performance of the trained models, the percentile-based error metric is used. The percentile-based metric includes the average and maximum of the stress field, and other representative percentiles like 50\%, 90\%, 97\%, and 99\%. The percentile-based error metric not only aids in understanding the model performance on field data. For example, it is possible certain frameworks are better at capturing the overall distribution, which indicates superior performance in the 50\% percentile metric, while others may be better at capturing higher stress points which will indicate superior performance in the 99\% metric.

\subsection{Base models Results}

As mentioned earlier, the base models are built with 100 training and validating samples and tested on 150 data points.

\subsubsection{PCA-GEBHM-PCA}

Table~\ref{tab:base_metrics} shows the percentile-based error metrics for testing data on non-rotated and rotated cases.
From the error metrics of the prediction by PCA-BHM-PCA, in both non-rotated and rotated cases, the average errors are less than 1\%, which affirms the model's ability to capture average stress fields.
For different percentiles of error metrics, the prediction error increases along with the increasing of percentiles, from $50^{th}$ percentile (median) to $100^{th}$ percentile (maximum). One possible reason for such a systemic increase could be the use of mean-squared error as a loss metric. The trained model seeks to capture the stress in the whole field at cost of larger stress values.
For the cases of non-rotated and rotated datasets, the error metrics are very close to each other across different percentile metrics. It was observed that the base model for the rotated case had a larger number of components than rotated cases owing to the higher complexity of rotated ellipse cases.

\begin{table*}[h]
\small
\sisetup{round-mode=places}
\centering
\caption{Testing data prediction metrics of two base modelings (100 samples) on non-rotated and rotated case \label{tab:base_metrics}}
\renewcommand{\arraystretch}{1.1}
\begin{tabular}{llllllll}
% \begin{tabular}{l
% l
% S[round-precision=3] 
% S[round-precision=3] 
% S[round-precision=3] 
% S[round-precision=3] 
% S[round-precision=3] 
% S[round-precision=3]}
\toprule
% {Models} & {Data type} & {Average} & {Maximum} & {50-th percentile} & {90-th percentile} & {97-th percentile} & {99-th percentile} \\
Model & Data type & Average & Maximum &  50-th percentile &  90-th percentile &  97-th percentile & 99-th percentile \\
\toprule
PCA-BHM-PCA & Non-rotated & \textbf{0.645}  & \textbf{12.5} & \textbf{0.633} & \textbf{2.30} & \textbf{3.69} & \textbf{5.55}  \\
PCA-NN-PCA & Non-rotated & 0.781 & 13.1 & 0.674 & 2.49 & 4.05 & 6.39  \\
\hline
PCA-BHM-PCA & Rotated & \textbf{0.716} & \textbf{10.2} & \textbf{0.674} & \textbf{2.31} & \textbf{3.74} & \textbf{5.14}  \\
PCA-NN-PCA & Rotated & 0.835 & 13.1 & 0.721 & 2.37 & 3.98 & 6.39  \\
\toprule
\end{tabular}
\end{table*}

\begin{figure}[h!]
\centering
\includegraphics[trim={0 0 0 0}, width=0.45\textwidth]{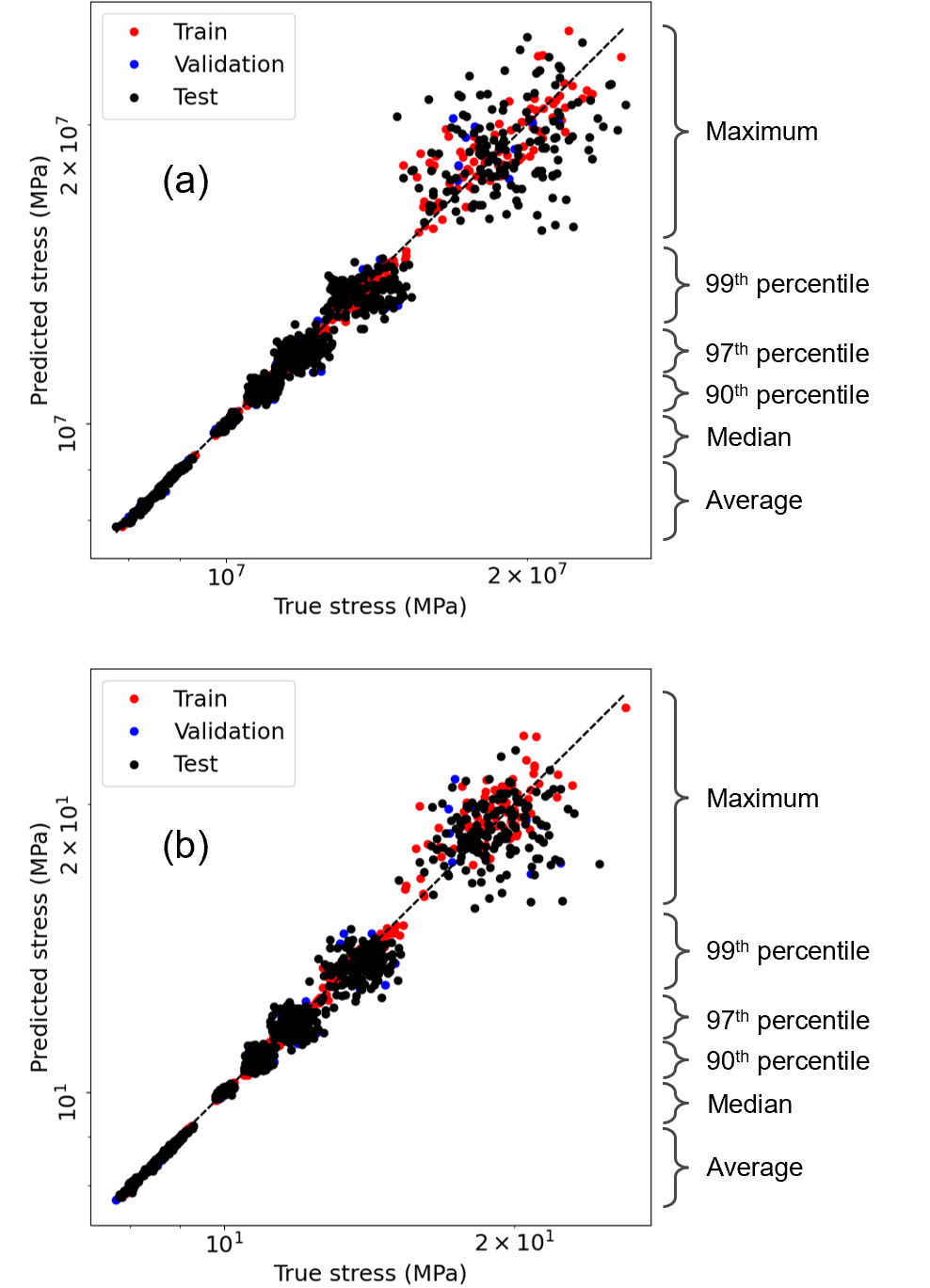}
\caption{Prediction error plots for training, validation, and testing data samples from the base model of PCA-BHM-PCA on (a) non-rotated and (b) rotated cases.}
\label{fig:N100_error_plot_bhm}
\end{figure}

\begin{figure}[h!]
\centering
\includegraphics[trim={0 0 0 0}, width=0.47\textwidth]{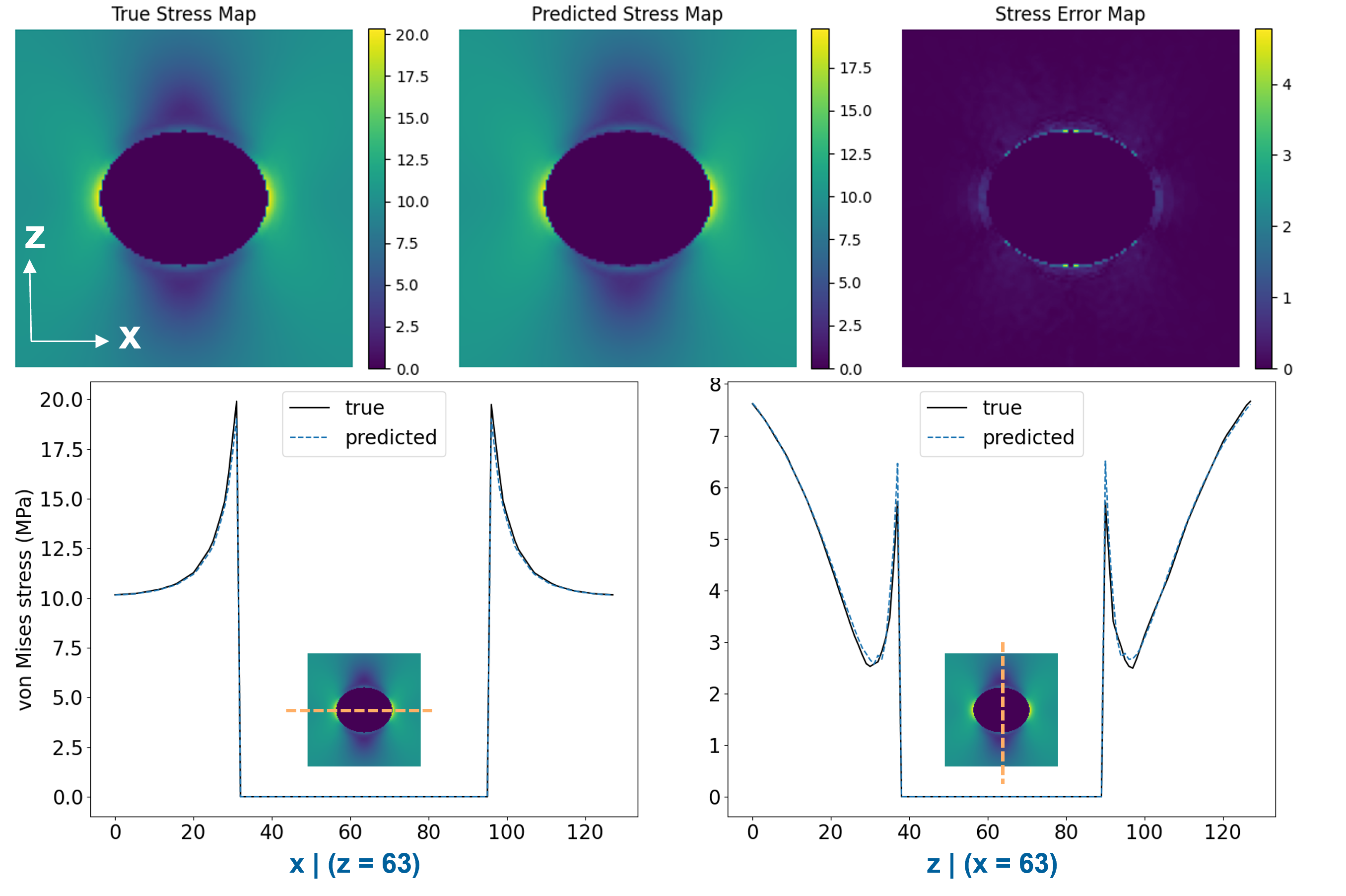}
\caption{Best prediction sample from PCA-BHM-PCA model on non-rotated case with stress map and stress at horizontal and vertical cross-sections.}
\label{fig:NonRotated_N100_best_bhm}
\end{figure}

\begin{figure}[h!]
\centering
\includegraphics[trim={0 0 0 0}, width=0.47\textwidth]{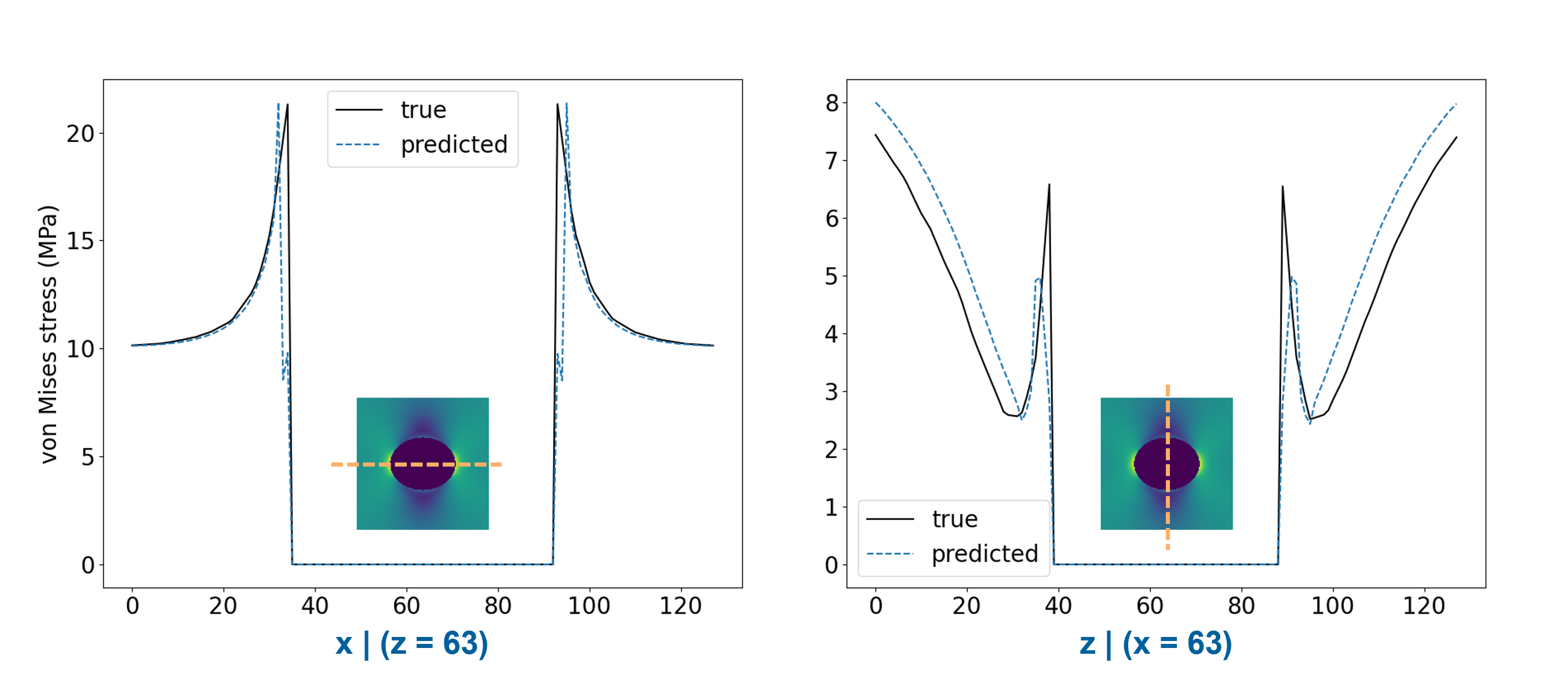}
\caption{Worst prediction sample from PCA-BHM-PCA model on non-rotated case with stress map and stress at horizontal and vertical cross-sections.}
\label{fig:NonRotated_N100_worst_bhm}
\end{figure}

\begin{figure}[h!]
\centering
\includegraphics[trim={0 0 0 0}, width=0.47\textwidth]{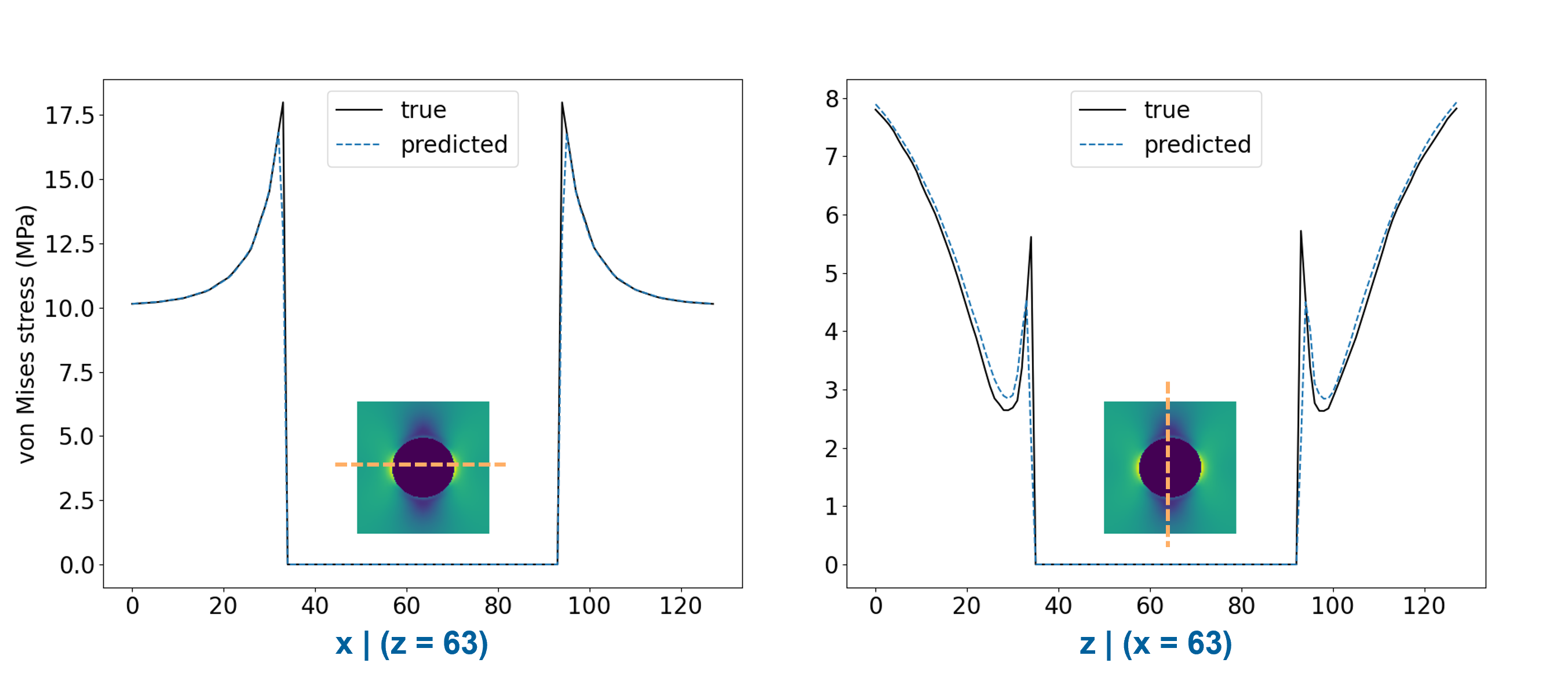}
\caption{Best prediction sample from PCA-BHM-PCA model on rotated case with stress map and stress at horizontal and vertical cross-sections.}
\label{fig:Rotated_N100_best_bhm}
\end{figure}

\begin{figure}[h!]
\centering
\includegraphics[trim={0 0 0 0}, width=0.47\textwidth]{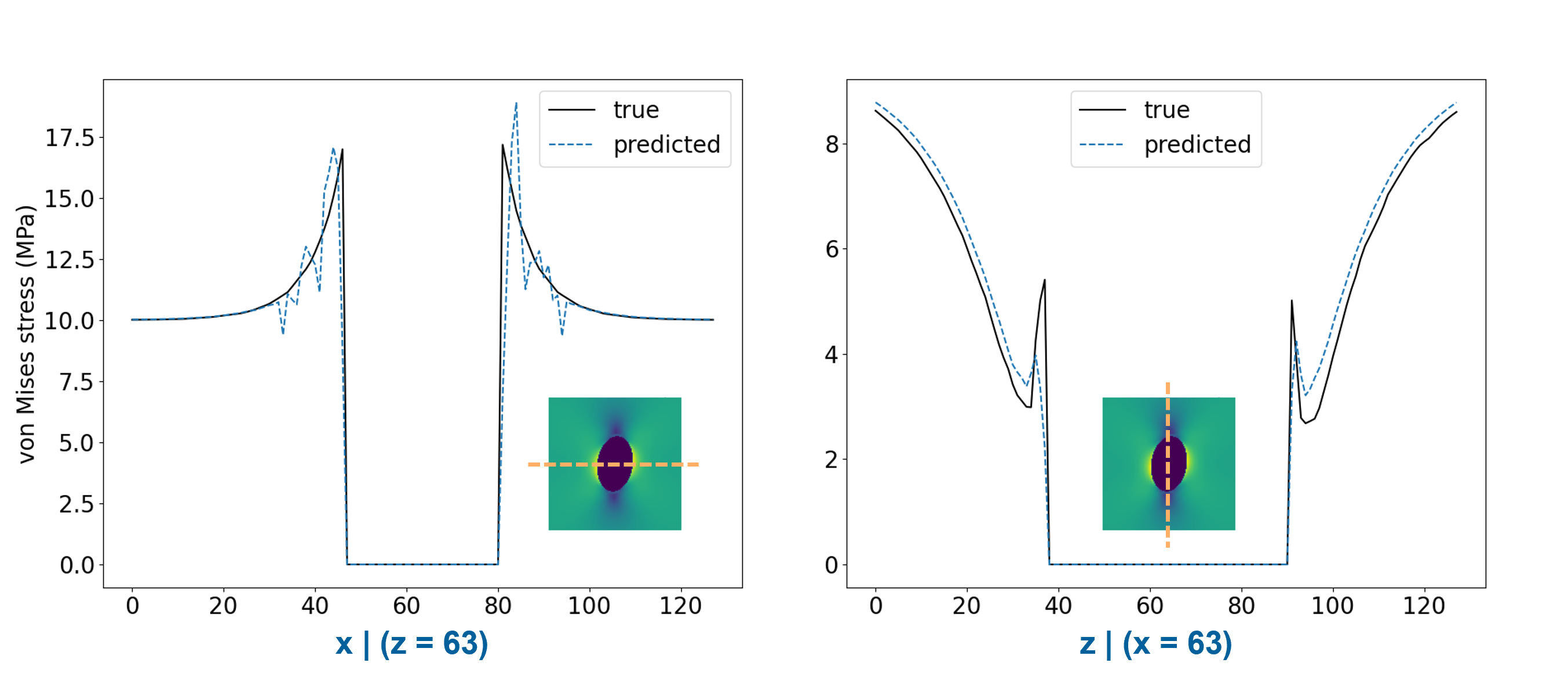}
\caption{Worst prediction sample from PCA-BHM-PCA model on rotated case with stress map and stress at horizontal and vertical cross-sections.}
\label{fig:Rotated_N100_worst_bhm}
\end{figure}

Figure \ref{fig:N100_error_plot_bhm} shows the 45-degree plots of the PCA-BHM-PCA predictions of different statistical measures of the stress field for all the training, validation, and testing datasets on (a) non-rotated and (b) rotated cases. These plots show a similar pattern of prediction accuracy on different percentiles as Table~\ref{tab:base_metrics}, where prediction error increases as the percentile of query increases. 
Figure~\ref{fig:NonRotated_N100_best_bhm} and Figure~\ref{fig:NonRotated_N100_worst_bhm} gives the stress map prediction for the test case with the highest and lowest prediction accuracy among the entire test dataset respectively on the non-rotated case. The best-predicted stress map shows that the trained model is able to capture the whole stress field. In the stress map cross-section, due to the post-process masking, the stress values are 0 at the void area, and the stress peak is well captured by the trained model.
Figure~\ref{fig:Rotated_N100_best_bhm} and Figure~\ref{fig:Rotated_N100_worst_bhm} give the stress map prediction from the test data with the highest and lowest prediction accuracy for the rotated case. As the larger rotation of the void leads to more complex stress fields, the best prediction sample is the shape with a small void rotation.
%

% % Nonrotated case
%
%

% % Rotated case
%

\subsubsection{PCA-NN-PCA}

As with the case of PCA-GEBHM-PCA, we see reduced accuracy for the high-stress regions (often around the void boundaries). This stress-dependent bias can be seen primarily in the 97th and 99th percentile regions and the global maximum values of the stress fields.  The underlying reason could be primarily the use of  mean-squared error in Deephyper's optimization framework. Alternative metrics that are weighted on the stress values could address some of the biases in the predictions. In this paper, however, we focus on using domain-agnostic metrics and highlighting stress reconstruction efficiencies. We reserve the exploration of tailored metrics and their performances for future studies. 

Between the rotated and non-rotated geometries, we notice that the increased complexity in the stress fields of the rotated geometry adversely affects the reconstruction efficiency. Overall, the best-performing rotated geometries are actually the ones with small rotation angles. On the other hand, the elongated ellipsoidal profiles with non-zero orientation with external force fields show degraded performance. In addition, the PCA reconstruction of rotated fields shows arc-like features in the residual fields (difference between true and predicted fields). Thus the residual fields show biases that are geometry-dependent. These errors signify the need for non-linear dimensionality reduction techniques for complex void geometries. 

% Plots - Stress profile for best/worst case

% placebolder for NN results

\begin{figure}[h!]
\centering
\includegraphics[trim={0 0 0 0}, width=0.45\textwidth]{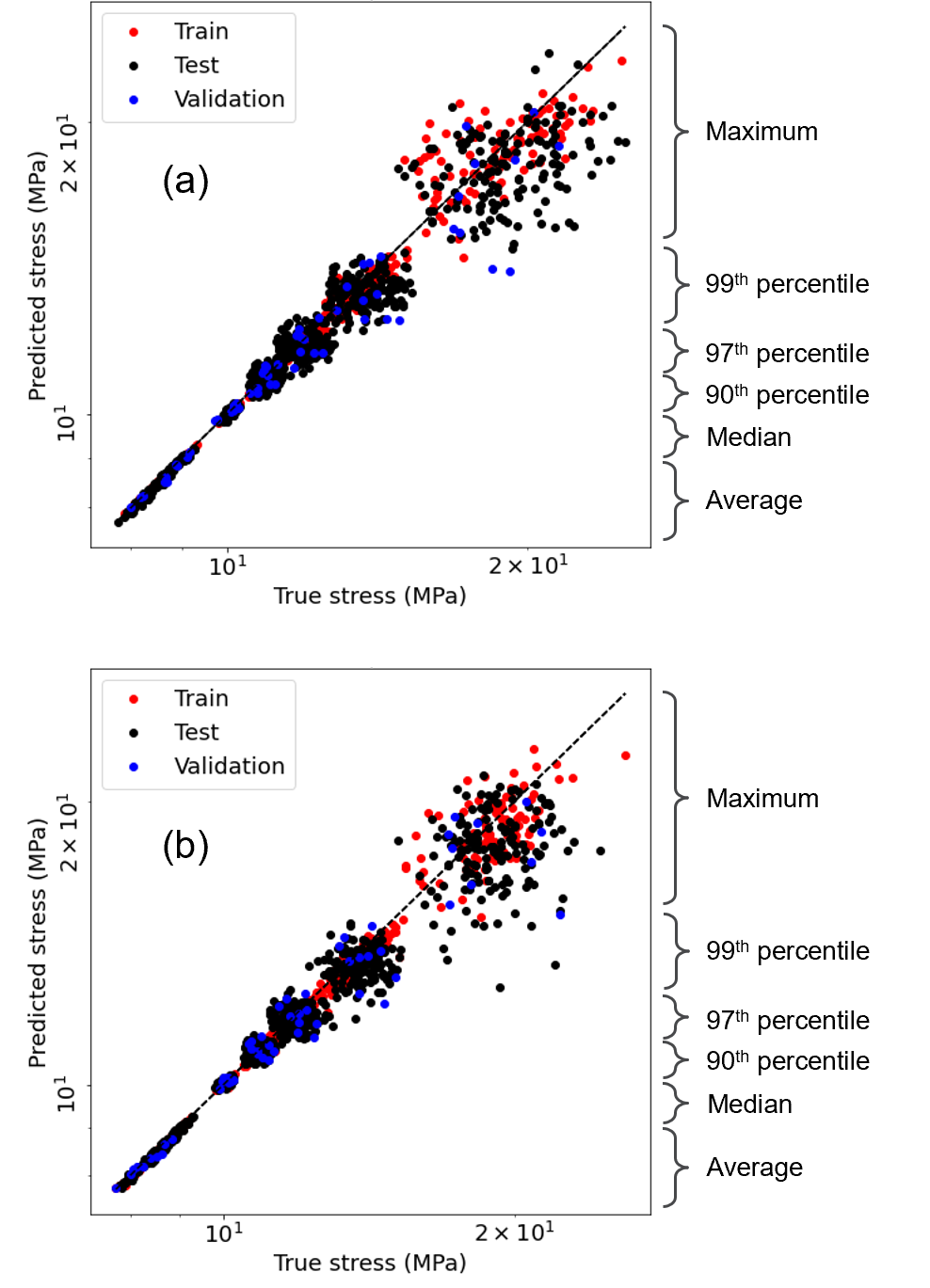}
\caption{Prediction error plots for training, validation and testing data samples from base model of PCA-NN-PCA on (a) non-rotated and (b) rotated cases.}
\label{fig:N100_error_plot_nn}
\end{figure}

% %
% \begin{figure}[h!]
% \centering
% \includegraphics[trim={0 0 0 0}, width=0.40\textwidth]{figures/NN/Rotated/All_stress_ploterror.png}
% \includegraphics[trim={0 0 0 0}, width=0.40\textwidth]{figures/NN/UnRotated/All_stress_ploterror.png}
% \caption{N100 error plot}
% \label{fig:N100_error_plot_nn}
% \end{figure}
% %

% % Nonrotated case
%
\begin{figure}[h!]
\centering
\includegraphics[trim={0 2.0 0 0}, width=0.47\textwidth]{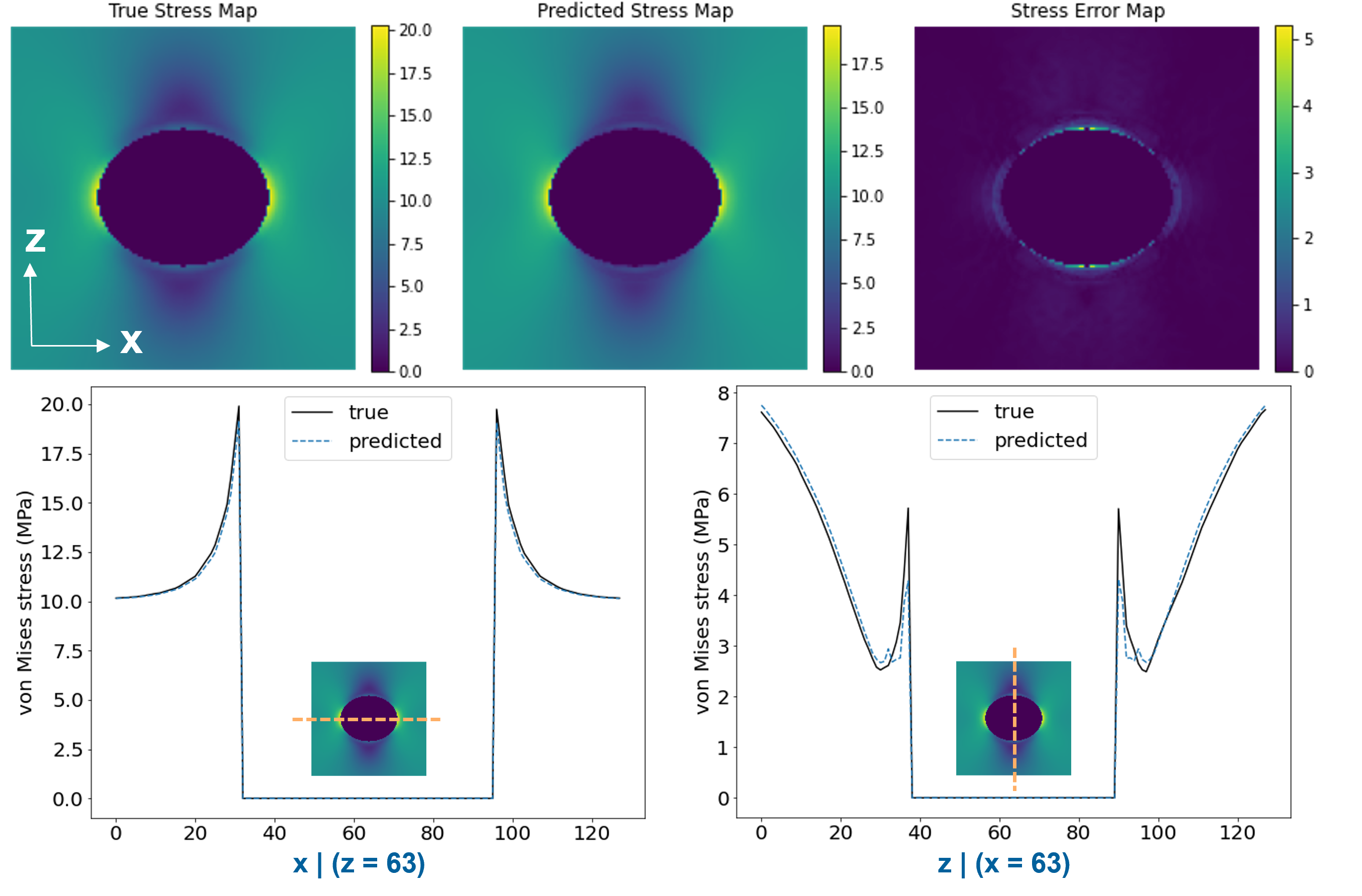}
\caption{Best prediction sample from PCA-NN-PCA model on non-rotated case with stress map and stress at horizontal and vertical cross-sections.}
\label{fig:NonRotated_N100_best_nn}
\end{figure}
\begin{figure}[h!]
\centering
\includegraphics[trim={0 0 0 0}, width=0.47\textwidth]{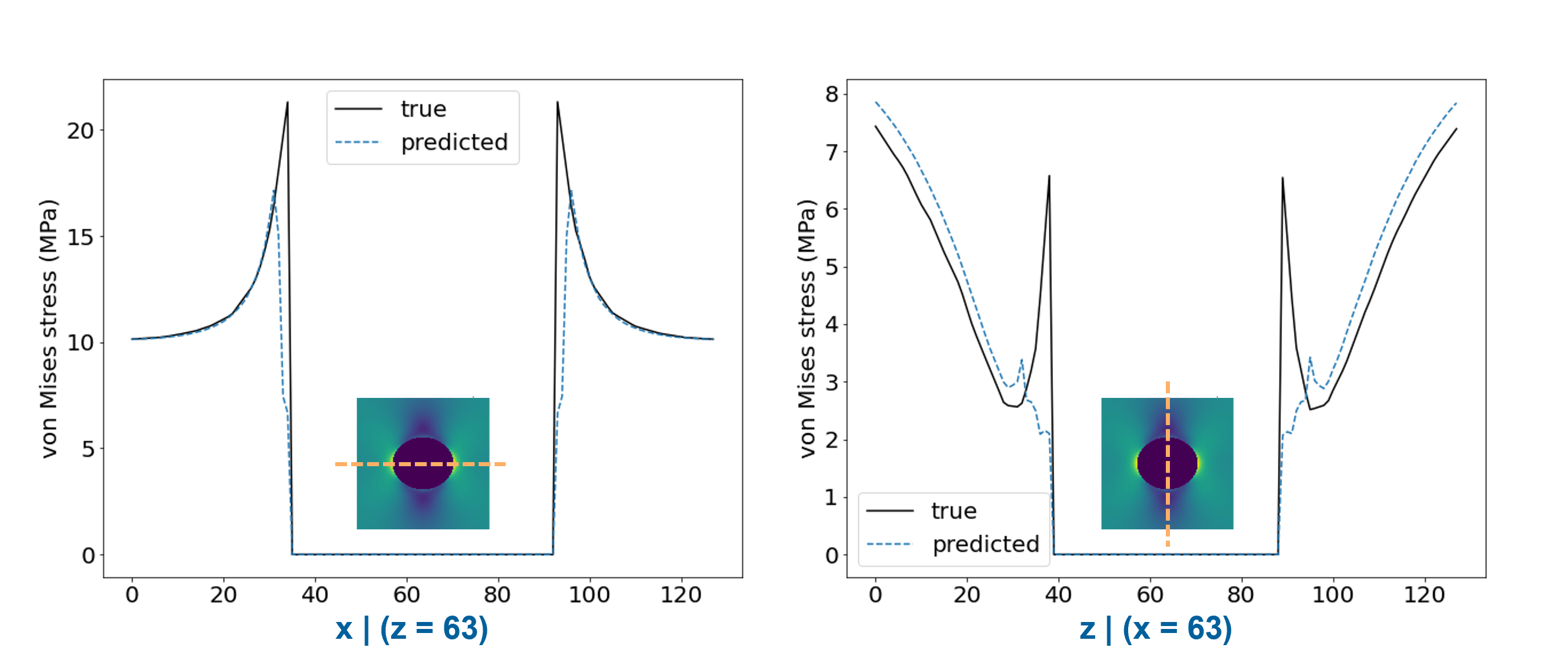}
\caption{Worst prediction sample from PCA-NN-PCA model on non-rotated case with stress map and stress at horizontal and vertical cross-sections.}
\label{fig:NonRotated_N100_worst_nn}
\end{figure}
%

% % Rotated case
%
\begin{figure}[h!]
\centering
\includegraphics[trim={0 0 0 0}, width=0.47\textwidth]{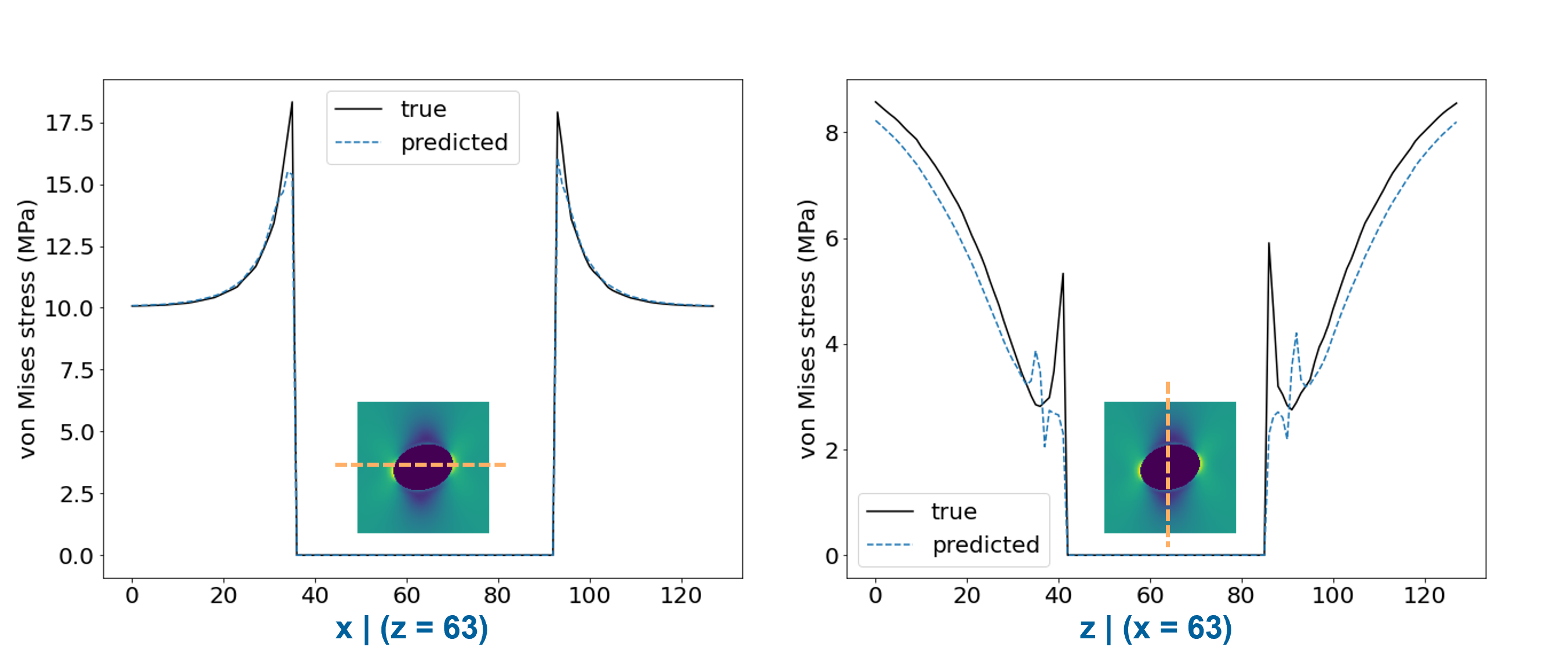}
\caption{Best prediction sample from PCA-NN-PCA model on rotated case with stress map and stress at horizontal and vertical cross-sections.}
\label{fig:Rotated_N100_best_nn}
\end{figure}
\begin{figure}[h!]
\centering
\includegraphics[trim={0 0 0 0}, width=0.47\textwidth]{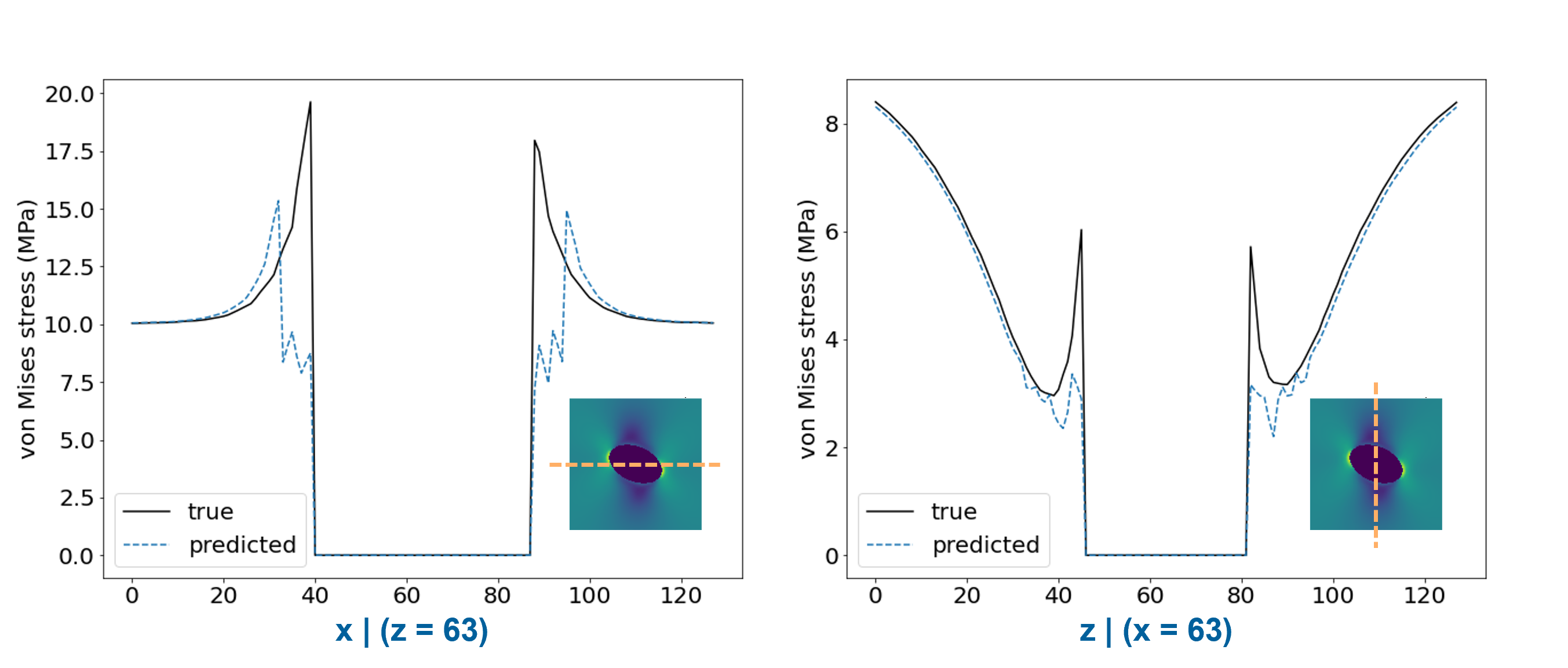}
\caption{Worst prediction sample from PCA-NN-PCA model on rotated case with stress map and stress at horizontal and vertical cross-sections.}
\label{fig:Rotated_N100_worst_nn}
\end{figure}
%

\begin{comment}

\begin{table*}[h]
\small
\centering
\caption{Training losses of two base modelings (100 samples) on Non-Rotated case ($\times 10^{11}$) \label{tab:loss_nonrotate}}
\renewcommand{\arraystretch}{1.3}
\begin{tabular}{llll}
\hline
Models & Training losses & Validation losses & Testing losses \\
\hline
BHM model & 0.7661940  & 4.03463234 & 3.83474112  \\
NN model & 0.0 & 0.0 & 0.0  \\
\hline
\end{tabular}
\end{table*}

\end{comment}

\begin{comment}

\begin{table*}[h]
\small
\centering
\caption{Training losses of two base modelings (100 samples) on Rotated case ($\times 10^{11}$) 
\label{tab:loss_rotate}}
\renewcommand{\arraystretch}{1.3}
\begin{tabular}{llll}
\hline
Models & Training losses & Validation losses & Testing losses \\
\hline
BHM model & 1.2267  & 5.7492 & 5.4473  \\
NN model & 0.0 & 0.0 & 0.0  \\
\hline
\end{tabular}
\end{table*}

\end{comment}

\begin{comment}
\begin{table*}[h]
\small
\centering
\caption{Testing data prediction metrics of two base modelings (100 samples) on Rotated case 
\label{tab:metrics_rotate}}
\renewcommand{\arraystretch}{1.3}
\begin{tabular}{lllllll}
\hline
Models & Average & Maximum & 50-th percentile & 90-th percentile & 97-th percentile & 99-th percentile \\
\hline
BHM dimensionality reduction & 0.0056505  & 0.099140 & 0.0068952 & 0.02316766 & 0.0370678 & 0.05389288  \\
Neural network modeling & 0.008351941 & 0.131136746 & 0.0072074871 & 0.023692110 & 0.039802075 & 0.063860894  \\
\hline
\end{tabular}
\end{table*}

\end{comment}

\subsection{Training Size Dependency}
% Convergence plots

With the successful development of the base models, a case study was conducted to understand the effect of data sizes on the proposed framework. 
To that end, additional four groups of datasets with varying numbers of training and validation samples are generated. The total numbers of training and validation data samples are 80, 60, 40, and 20 in these groups, and 90\% samples for training and 10\% for validation. The testing data sampled is kept the same as the previous section results with 150 samples for both non-rotated and rotated cases. Table~\ref{tab:sizes_metrics_nonrotate} and Table~\ref{tab:sizes_metrics_rotate} shows the testing data prediction metrics for different sizes of training data on non-rotated and rotated case, respectively. As expected, a drop is observed in error metrics as the training data size increases. Similar to the base models, across all sizes of training samples tested here, both models give the best prediction on average stress, and the prediction error increase along the increasing of query percentile. 

\begin{comment}
\begin{table*}[h]
\small
\centering
\caption{Training losses across different sizes of training data on Non-Rotated case ($\times 10^{11}$)
\label{tab:sizes_losses_nonrotate}}
\renewcommand{\arraystretch}{1.3}
\begin{tabular}{lllll}
\hline
Samples & Models & Training losses & Validation losses & Testing losses \\
\hline
80 & BHM model & 0.742313  & 4.759498 & 4.655959 \\
80 & NN modeling & 0.0 & 0.0 & 0.0 \\
60 & BHM model & 0.714389  & 4.842250 & 5.185138 \\
60 & NN modeling & 0.0 & 0.0 & 0.0 \\
40 & BHM model & 0.494808  & 7.399856 & 5.924811 \\
40 & NN modeling & 0.0 & 0.0 & 0.0  \\
20 & BHM model & 0.550712 & 6.230323 & 9.135567 \\
20 & NN modeling & 0.0 & 0.0 & 0.0 \\
\hline
\end{tabular}
\end{table*}

\end{comment}

\begin{table*}[h]
\small
\sisetup{round-mode=places}
\centering
\caption{Testing data prediction metrics across different sizes of training data on non-rotated case \label{tab:sizes_metrics_nonrotate}}
\renewcommand{\arraystretch}{1.3}
\begin{tabular}{llllllll}
% \begin{tabular}{l 
% l 
% S[round-precision=3] 
% S[round-precision=3] 
% S[round-precision=3] 
% S[round-precision=3] 
% S[round-precision=3]
% S[round-precision=3]}
\toprule
% {\# Samples} & {Model} & {Average} & {Maximum} & {50-th percentile} & {90-th percentile} & {97-th percentile} & {99-th percentile} \\
\# Samples & Model & Average & Maximum & 50-th percentile & 90-th percentile & 97-th percentile & 99-th percentile \\
\toprule
80 & PCA-BHM-PCA &  \textbf{0.716}  & \textbf{12.3} & \textbf{0.660} & \textbf{2.43} & \textbf{3.83} & \textbf{5.95}  \\
 & PCA-NN-PCA & 0.818 & 12.9 & 0.701 & 2.63 & 4.35 & 7.48  \\
\hline
60 & PCA-BHM-PCA & \textbf{0.742}  & 13.6 & \textbf{0.673} & \textbf{2.44} & \textbf{3.81} & \textbf{6.14}  \\
 & PCA-NN-PCA & 0.878 & \textbf{12.9} & 0.674 & 2.52 & 4.23 & 6.65  \\
\hline
40 & PCA-BHM-PCA & \textbf{0.741}  & 14.1 & 0.753 & 2.64 & 4.05 & \textbf{6.35}  \\
& PCA-NN-PCA &  0.902 & \textbf{13.8} & \textbf{0.704} & \textbf{2.48} & \textbf{4.01} & 6.85  \\
\hline
20 & PCA-BHM-PCA & \textbf{0.836}  & \textbf{15.4} & \textbf{0.666} & 2.70 & 4.19 & \textbf{6.47}  \\
& PCA-NN-PCA & 1.25 & 18.9 & 0.926 &  \textbf{2.28} & \textbf{3.89} & 7.95 \\
\toprule
\end{tabular}
\end{table*}

\begin{comment}
\begin{table*}[h]
\small
\centering
\caption{Training losses across different sizes of training data on Rotated case ($\times 10^{11}$)
\label{tab:sizes_losses_rotate}}
\renewcommand{\arraystretch}{1.3}
\begin{tabular}{lllll}
\hline
Samples & Models & Training losses & Validation losses & Testing losses \\
\hline
80 & BHM model & 1.54485243  & 5.241941556 & 5.9026159529 \\
80 & NN modeling & 0.0 & 0.0 & 0.0 \\
60 & BHM model & 0.756806653  & 6.8668891592 & 6.39505519969 \\
60 & NN modeling & 0.0 & 0.0 & 0.0 \\
40 & BHM model & 0.790575022  & 7.3660886124 & 8.21884114780 \\
40 & NN modeling & 0.0 & 0.0 & 0.0  \\
20 & BHM model & 1.107143428  & 13.834269065 & 11.728540673 \\
20 & NN modeling & 0.0 & 0.0 & 0.0 \\
\hline
\end{tabular}
\end{table*}

\end{comment}

\begin{table*}[h]
\small
\sisetup{round-mode=places}
\centering
\caption{Testing data prediction metrics across different sizes of training data on rotated case \label{tab:sizes_metrics_rotate}}
\renewcommand{\arraystretch}{1.3}
\begin{tabular}{llllllll}
% \begin{tabular}{l 
% l 
% S[round-precision=3] 
% S[round-precision=3] 
% S[round-precision=3] 
% S[round-precision=3] 
% S[round-precision=3]
% S[round-precision=3]}
\toprule
% {\# Samples} & {Model} & {Average} & {Maximum} & {50-th percentile} & {90-th percentile} & {97-th percentile} & {99-th percentile} \\
\# Samples & Model & Average & Maximum & 50-th percentile & 90-th percentile & 97-th percentile & 99-th percentile \\
\toprule
80 & PCA-BHM-PCA & \textbf{0.725}  & \textbf{11.1} & \textbf{0.693} & \textbf{2.27} & \textbf{3.69} & \textbf{5.42} \\
& PCA-NN-PCA & 0.866 & 13.1 & 0.792 & 2.41 & 4.09 & 6.31 \\
\hline
60 & PCA-BHM-PCA & \textbf{0.666} & \textbf{10.9} & \textbf{0.624} & \textbf{2.14} & \textbf{3.45} & \textbf{5.31} \\
& PCA-NN-PCA & 0.893 & 15.0 & 0.726 & 2.20 & 3.70 & 6.51 \\
\hline
40 & PCA-BHM-PCA & \textbf{0.747} & \textbf{12.8} & \textbf{0.669} & 2.30 & 3.73 & \textbf{5.21}  \\
& PCA-NN-PCA & 1.07 & 17.1 & 0.784 & \textbf{2.18} & \textbf{3.72} & 7.35 \\
\hline
20 & PCA-BHM-PCA & \textbf{1.01} & \textbf{15.5} & \textbf{0.826} & \textbf{2.19} & \textbf{3.37} & \textbf{5.83}  \\
& PCA-NN-PCA & 1.25 & 18.9 & 0.926 & 2.28 & 3.89 & 7.95 \\
\toprule
\end{tabular}
\end{table*}

%%%%%%%%%%%%%%%%%%%%%%%%%%%%%%%%%%%%%%%%%%%%%%%%%%%%%%%%%%%%%%%%%%%%%%
\clearpage
\section{Conclusions} \label{sec:conclusions}

The work done in the paper set out to demonstrate a scalable modeling framework for high-dimensional data leveraging dimensionality reduction (through PCA) and surrogate modeling (through GEBHM and Deephyper NN). The proposed framework was tested on a representative ellipsoidal void problem with and without rotation.

The framework shows satisfactory performance with limited data and stands as a promising approach to modeling high-dimensional spare data problems. A visible trend of better performance on the low percentile metric was observed compared to the higher percentile metric. This could be possibly attributed to the use of mean-squared error as a loss metric. Further investigation might be needed to augment this aspect of the framework.

Additionally, a comprehensive study was conducted to observe the performance of the models across different datasets. This is particularly important for problems involving expensive machining or high-fidelity simulations. We see an expected trend of increasing average error with reducing the number of training points. Hence, depending on the error budget and available training points, extensive model investigations and parameter searches are necessary. 

In this study, we have only worked with a relatively simple ellipsoidal void problem. With increased complexity in the void geometries, stress inference is bound to have higher prediction biases. Such studies might require higher-order non-linear dimensionality reduction methods, coupled with more extensive latent space mapping searches. This is another aspect of the work that warrants further investigation.

%Stress recovery can show geometry-based or stress value-based biases. This can be utilized in training the models. 

%%%%%%%%%%%%%%%%%%%%%%%%%%%%%%%%%%%%%%%%%%%%%%%%%%%%%%%%%%%%%%%%%%%%%%
% Here's where you specify the bibliography style file.
% The full file name for the bibliography style file 
% used for an ASME paper is asmems4.bst.
%%%%%%%%%%%%%%%%%%%%%%%%%%%%%%%%%%%%%%%%%%%%%%%%%%%%%%%%%%%%%%%%%%%%%%
\clearpage
\section*{Acknowledgments}
This material is based upon work supported by the U.S. Department of Energy’s Office of Energy Efficiency and Renewable Energy (EERE) under the Advanced Manufacturing Office, Award Number DE-AC0206H11357.
The views expressed herein do not necessarily represent the views of the U.S. Department of Energy or the United States Government. 
Work at Argonne National Laboratory was supported by the U.S. Department of Energy, Office of High Energy Physics. Argonne, a U.S. Department of Energy Office of Science Laboratory, is operated by UChicago Argonne LLC under contract no. DE-AC02-06CH11357. 
This manuscript has been authored by UT-Battelle, LLC under
Contract No. DE-AC05-00OR22725 with the US Department of Energy.
The United States Government retains and the publisher, by accepting
the article for publication, acknowledges that the United States
Government retains a non-exclusive, paid-up, irrevocable, world-wide
license to publish or reproduce the published form of this manuscript
or allow others to do so, for United States Government purposes.
The Department of Energy will provide public access to these results of
federally sponsored research in accordance with the DOE Public Access Plan (\url{http://energy.gov/downloads/doe-public-access-plan}).
Part of the analysis here is carried out on Swing, a GPU system at the Laboratory Computing Resource Center (LCRC) of Argonne National Laboratory.
We would also like to thank Dr. Aymeric Moinet, at General Electric Research, for his insights into the ellipsoidal void problem.
%%%%%%%%%%%%%%%%%%%%%%%%%%%%%%%%%%%%%%%%%%%%%%%%%%%%%%%%%%%%%%%%%%%%%%
% The bibliography is stored in an external database file
% in the BibTeX format (file_name.bib).  The bibliography is
% created by the following command and it will appear in this
% position in the document. You may, of course, create your
% own bibliography by using thebibliography environment as in
%
% \begin{thebibliography}{12}
% ...
% \bibitem{itemreference} D. E. Knudsen.
% {\em 1966 World Bnus Almanac.}
% {Permafrost Press, Novosibirsk.}
% ...
% \end{thebibliography}

% Here's where you specify the bibliography database file.
% The full file name of the bibliography database for this
% article is asme2e.bib. The name for your database is up
% to you.

%Bibliography
\clearpage
\bibliographystyle{unsrt}  
\bibliography{references}  

\end{document}